\documentclass[RNAAS]{aastex63}

\usepackage{color}
\newcommand{\target}{CWISE~J203546.35-493611.0}
\newcommand{\starget}{CWISE~J2035-4936}

\begin{document}

\title{Backyard Worlds: Planet 9 discovery of an unusual low-mass companion to an M Dwarf at 80 pc}

\author[0000-0003-4083-9962]{Austin Rothermich}
\affiliation{Physics Department, University Of Central Florida, 4000 Central Florida Boulevard, Orlando, FL 32816, USA}

\author[0000-0002-6294-5937]{Adam C. Schneider}
\affiliation{US Naval Observatory, Flagstaff Station, P.O. Box 1149, Flagstaff, AZ 86002, USA}
\affiliation{Department of Physics and Astronomy, George Mason University, MS3F3, 4400 University Drive, Fairfax, VA 22030, USA}

\author[0000-0001-6251-0573]{Jacqueline K. Faherty}
\affiliation{Department of Astrophysics, American Museum of Natural History, Central Park West at 79th Street, NY 10024, USA}

\author[0000-0003-0580-7244]{Katelyn Allers}
\affiliation{Bucknell University; Department of Physics and Astronomy; Lewisburg, PA 17837, USA}

\author[0000-0001-8170-7072]{Daniella Bardalez-Gagliuffi}
\affiliation{Department of Astrophysics, American Museum of Natural History, Central Park West at 79th Street, NY 10024, USA}

\author[0000-0002-1125-7384]{Aaron M. Meisner}
\affiliation{NSF's National Optical-Infrared Astronomy Research Laboratory, 950 N. Cherry Ave., Tucson, AZ 85719, USA}

\author[0000-0002-2387-5489]{Marc Kuchner}
\affiliation{NASA Goddard Space Flight Center, Exoplanets and Stellar Astrophysics Laboratory, Code 667, Greenbelt, MD 20771, USA}

\author[0000-0003-4269-260X]{J. Davy Kirkpatrick}
\affiliation{IPAC, Mail Code 100-22, Caltech, 1200 E. California Blvd., Pasadena, CA 91125, USA}

\author[0000-0001-7896-5791]{Dan Caselden}
\affiliation{Gigamon Applied Threat Research, 619 Western Avenue, Suite 200, Seattle, WA 98104, USA}

\author{Paul Beaulieu}
\affiliation{Backyard Worlds: Planet 9}


\begin{abstract}
    We present the discovery of \target, a peculiar M8 companion to the M4.5 star APMPM J2036-4936 discovered through the citizen science project Backyard Worlds: Planet 9. Given \target's proper motion ($\mu_{\alpha}$, $\mu_{\delta}$) = ($-$126$\pm$22, $-$478$\pm$23) and angular separation of 34.2$''$ from APMPM 2036-4936, we calculate a chance alignment probability of $1.15 \times 10^{-6}$. Both stars in this system appear to be underluminous, and the spectrum obtained for \target\ shows a triangular H band. Further study of this system is warranted to understand these peculiarities.
\end{abstract}

\section{Introduction}
Low-mass companions to known stars provide an excellent opportunity to better understand the formation and evolution of these objects. Identifying co-moving companions has been made easier through the use of all sky, multi-epoch surveys such as the {\it Wide-field Infrared Survey Explorer} ({\it WISE}; \citealt{2010AJ....140.1868W}). Backyard Worlds: Planet 9 \citep{2017ApJ...841L..19K} is a citizen science project where volunteers examine {\it WISE} images to identify high proper motion objects. Here we report the discovery of \target\ (hereafter \starget) an object at the M/L spectral type boundary co-moving with a known M4.5 star found through the Backyard Worlds project. 

\section{Discovery of Companion}
The primary of this system, APMPM J2036-4936, was discovered and classified via an optical spectrum as M4.5 in \cite{2005A&A...440.1061L}. Those authors estimated a distance of 165.8 pc to APMPM J2036-4936 using photometric-spectral type relations, resulting in a tangential velocity estimate of 335 km s$^{-1}$. Using the parallax measurement for this object from Gaia DR2 \citep{2018AandA...616A...1G}, we find a new distance of 81.5$\pm$1.5 pc, resulting in a tangential velocity of 162 km s$^{-1}$. Using this new distance along with the Gaia DR2 G magnitude for APMPM J2036-4936 of G=18.080$\pm$0.002 mag, we calculate an absolute G magnitude of 13.52 mag. Using the relations described in \cite{2019AJ....157..231K}, this absolute magnitude plus its colors in Gaia DR2 (BP-RP=3.738 mag, BP-G=2.313 mag, G-RP=1.425 mag) place APMPM J2036-4936 as M7. There is no clear explanation for why the Gaia photometry for APMPM J2036-4936 appears underluminous, suggesting an M7 spectral type using the Gaia color-magnitude relations despite it being classified via its optical spectrum as a normal M4.5. Either this source has highly unusual properties or the wrong target was observed in \cite{2005A&A...440.1061L}, possibly due to its large proper motion.  

The companion, \starget\, was independently found by two citizen scientists: Paul Beaulieu and Austin Rothermich. It was identified by visually inspecting APMPM J2036-4936 using the WiseView tool \citep{2018ascl.soft06004C}. Given the existing VHS photometry \citep{2013Msngr.154...35M} (J=16.810 $\pm$ 0.010 mag, H=16.444 $\pm$ 0.015 mag, K=16.028 $\pm$ 0.0229 mag) and CatWISE 2020 photometry \citep{2020ApJS..247...69E} (W1=15.742 $\pm$ 0.023 mag, W2=15.455 $\pm$ 0.041 mag) and distance of APMPM J2036-4936, we estimated a spectral type between L2 and L5. The CatWISE 2020 proper motion for \starget \ is  ($\mu_{\alpha}$, $\mu_{\delta}$) = ($-$126$\pm$22, $-$478$\pm$23) mas/yr, compared to ($\mu_{\alpha}$, $\mu_{\delta}$) = ($-$82$\pm$0.286, $-$411$\pm$0.279) mas/yr for APMPM J2036-4936 from Gaia DR2.   

\section{Spectroscopic Observations}
\begin{figure}
    \centering
    \includegraphics[width=1\textwidth]{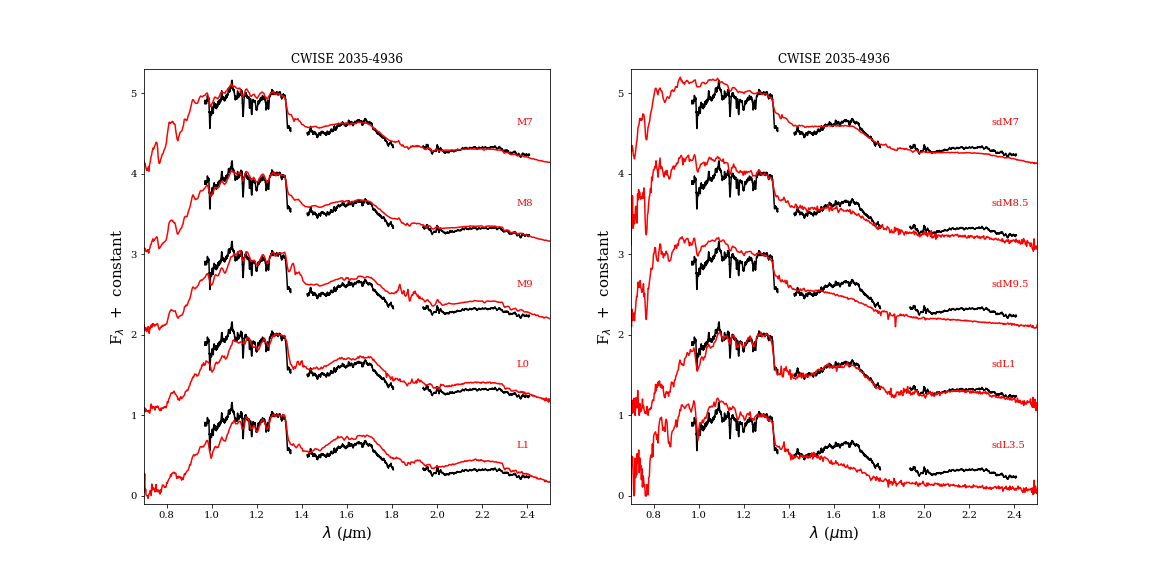}
    \caption{Left: spectrum of \starget\ (black) compared to the spectral standards VB 8 (M7)\citep{2010ApJS..190..100K}, VB 10 (M8)\citep{2010ApJS..190..100K}, LHS 2924 (M9)\citep{2010ApJS..190..100K}, 2MASS 0345+2540 (L0)\citep{2010ApJS..190..100K}, and 2MASS 2130-0845 (L1)\citep{2010ApJS..190..100K}. Right: \starget\ compared to LHS 377 (sdM7)\citep{1997AJ....113..806G,2019AJ....158..182G}, 2MASS 0142+0523 (sdM8.5)\citep{2007ApJ...657..494B,2019AJ....158..182G}, SSSPM 1013 (sdM9.5)\citep{2014ApJ...783..122K}, and 2MASS 1756+2815 (sdL1)\citep{2010ApJS..190..100K,2019AJ....158..182G}, and SDSS J1256-0224 (sdL3.5)\citep{2009ApJ...697..148B,2019AJ....158..182G}  All spectra were normalized between 1.27-1.29 microns and separated by a constant.}
    \label{fig:spectrum}
\end{figure}

We obtained a near-infrared (0.97--2.41 $\mu m$) spectrum of \starget\ on UT 2019-06-19 using TripleSpec \citep{2004SPIE.5492.1295W} on the Southern Astrophysical Research (SOAR) Telescope.  Using an ABBA nod pattern, we observed a total of 12 exposures of 180 seconds each.  Our science observations were taken at an airmass of 1.11"--1.17" under conditions of scattered high cirrus.  Immediately following our observations of \starget\, we observed the A0 star, HD~198546 ($8 \times 5$ second exposures at an airmass of 1.16") for telluric calibration.  We reduced our data using a modified version of Spextool \citep{2004PASP..116..362C}, including a correction for telluric absorption following the method described in \citet{2003PASP..115..389V}. The spectrum of \starget\ compared to various spectral standards is shown in Figure \ref{fig:spectrum}. 

\section{Discussion}
The spectrum obtained for \starget\ does not fully match any of the spectral standards. When looking at the J band portion of the spectrum however, the closest fit is with that of VB 8 (M8), although there are still a few minor features which do not fully match. The K band shows some slight irregularities when compared with VB 8, such as slightly lower flux near the blue end. The H band of \starget\ however is quite peculiar, appearing to have more of a triangular shape than the H band of a standard M8, a feature that has been seen in metal low, sub-dwarf stars \citep{Aganze_2016}. This led us to compare \starget\ to sub-dwarf standards as shown in the right panel of Figure \ref{fig:spectrum}, which yielded no good fits, though the sdL1 was the closest match. We find no indication that the odd features seen in \starget\ are due to unresolved binarity following the spectral binary template comparison of \cite{2010ApJ...710.1142B} and \cite{2014ApJ...794..143B}. Because the best fit with the normal standards is with that of the M8, we assign \starget\ a near-infrared spectral type of M8 pec. We find that using a very conservative distance range of 67 to 349 pc, found by using the Gaia absolute magnitude relations in \cite{2019AJ....157..231K} for a spectral type range of M7-L2 and the existing Gaia photometry, 166 objects in Gaia EDR3 \cite{2020arXiv201201533G} match with $\mu_{\alpha}$ and $\mu_{\delta}$ both within $\pm$70 mas/yr of the corresponding \starget \ values (70 mas/yr $\approx$ 3$\times$max[$\sigma_{\mu_{\alpha}}$, $\sigma_{\mu_{\delta}}$]). Given the angular separation of 34.2$''$, this yields a chance alignment probability of $1.15 \times 10^{-6}$. The angular separation corresponds to a projected physical separation of $\sim$2,790 AU. Like APMPM J2036-4936, \starget\ also appears to be underluminous, as using our earlier  estimates it appeared to be within the range of L2-L5. Due to this apparent underluminosity, the high tangential velocity calculated above, and the triangular H band, we do not rule out the possibility of low metallicity in \starget. Further study of this system is required to understand its peculiarities.    

\section{Acknowledgements}
This work has made use of data from the European Space Agency (ESA) mission
{\it Gaia} (\url{https://www.cosmos.esa.int/gaia}), processed by the {\it Gaia}
Data Processing and Analysis Consortium (DPAC,
\url{https://www.cosmos.esa.int/web/gaia/dpac/consortium}). Funding for the DPAC
has been provided by national institutions, in particular the institutions
participating in the {\it Gaia} Multilateral Agreement.
\bibliographystyle{aasjournal}
\bibliography{byw.bib}

\end{document}